\documentclass[prl,twocolumn,amsmath,amssymb]{revtex4} 
\usepackage{graphicx} 
\usepackage{amsmath}
\usepackage{amsfonts,amsbsy}
\usepackage{amssymb}

\def\gsim{ \,\, \vcenter{\hbox{$\buildrel{\displaystyle >}\over\sim$}}
 \,\,}
\def\lsim{ \,\, \vcenter{\hbox{$\buildrel{\displaystyle <}\over\sim$}}
 \,\,}
\def\be{\begin{equation}}
\def\ee{\end{equation}}
\def\bea{\begin{eqnarray}}
\def\eea{\end{eqnarray}}
\def\tr{{\rm tr}\,}
\newcommand{\ket}[1]{\left| #1 \right>} 
\newcommand{\sgn}{\mathop{\mathrm{sgn}}}

\begin{document}

\title{\bf Magnetic flux loop in high-energy heavy-ion collisions}

\preprint{RBRC-xyz}

\author{Adrian Dumitru$^{a,b,c}$, Yasushi Nara$^{d}$ and Elena Petreska$^{b,c}$}
\affiliation{
$^a$ RIKEN BNL Research Center, Brookhaven National
  Laboratory, Upton, NY 11973, USA\\
$^b$ Department of Natural Sciences, Baruch College, CUNY,
17 Lexington Avenue, New York, NY 10010, USA\\
$^c$ The Graduate School and University Center, The City
  University of New York, 365 Fifth Avenue, New York, NY 10016, USA\\
$^d$ Akita International University, Yuwa, Akita-city 010-1292,
Japan\\
}

\begin{abstract}
We consider the expectation value of a chromo-magnetic flux loop in the
immediate forward light cone of collisions of heavy nuclei at high
energies. Such collisions are characterized by a non-linear scale
$Q_s$ where color fields become strong. We find that loops of area
greater than $\sim 1.5/Q_s^2$ exhibit area law behavior, which
determines the scale of elementary flux excitations (``vortices''). We
also estimate the magnetic string tension, $\sigma_M \simeq
0.12 \, Q_s^2$. By the time $t\sim 1/Q_s$ even small loops satisfy
area law scaling. We describe corrections to the propagator of semi-hard
particles at very early times in the background of fluctuating magnetic
fields.
\end{abstract}

\maketitle

Collisions of heavy ions at high energies provide opportunity to study
non-linear dynamics of strong QCD color fields~\cite{Mueller:1999wm}.
The field of a very dense system of color charges at rapidities far
from the source is determined by the classical Yang-Mills equations
with a recoilless current along the light cone~\cite{MV}. It consists
of gluons characterized by a transverse momentum $p_T$ on the order
of the density of valence charges per unit transverse area
$Q_s^2$~\cite{JalilianMarian:1996xn}; this {\em saturation momentum}
scale separates the regime of non-linear color field interactions at
$p_T\lsim Q_s$ or distances $r\gsim 1/Q_s$ from the perturbative
regime at $p_T\gg Q_s$. Near the center of a large nucleus this scale
is expected to exceed $\sim 1.5$~GeV at BNL-RHIC or CERN-LHC collider
energies, for a probe in the adjoint representation of the color gauge
group. The classical field solution provides the leading contribution
to an expansion in terms of the coupling and of the inverse saturation
momentum.

The soft field produced in a collision of two nuclei is then a
solution of the Yang-Mills equations satisfying appropriate matching
conditions on the light cone~\cite{Kovner:1995ja}. Most interestingly,
right after the impact strong longitudinal chromo-magnetic fields
$B_z\sim 1/g$ develop due to the fact that the individual projectile
and target fields do not commute~\cite{Fries:2006pv,TL_LM}. They
fluctuate according to the random local color charge
densities of the valence sources. In this Letter we show that magnetic
loops $W_M$ exhibit area law behavior, and we compute the magnetic
string tension. Furthermore, we argue that at length scales $\sim
1/Q_s$ the field configurations might be viewed as uncorrelated Z(N)
vortices. At finite times $\sim 1/Q_s$ after the collision area law
behavior is observed even for rather small Wilson loops. Finally, we
sketch how the background of magnetic fields affects propagation of
semi-hard particles with transverse momenta somewhat above $Q_s$.

Consider a spatial Wilson loop with radius $R$ in the plane
transverse to the beams,
\bea 
M(R) &=& {\cal P}\exp\left(
ig \int\limits_{-\pi}^\pi
d\theta \; \frac{\partial x^i}{\partial\theta} A^i\right) \nonumber\\
W_M(R) &=& \frac{1}{N_c} \left< \tr M(R)\right>~,
\label{eq:W_M}
\eea
where $x=R(\cos\theta,\sin\theta)$, and path ordering is with respect
to the angle $\theta$; in numerical lattice simulations it is more
convenient to employ a square loop. We compare, also, to the
expectation value of the Z($N_c$) part of the loop; for a magnetic field
configuration corresponding simply to a superposition of independent
vortices the loop should equal $\exp (2 \pi i \, n/N_c)$, with $n$ the
total vortex charge piercing the loop. Thus, for two colors we compute
\be  \label{eq:W_M_Z2}
W_M^{Z(2)}(R) = \left< \sgn \tr M(R)\right>
\ee
where $\sgn()$ denotes the sign function. Comparing~(\ref{eq:W_M})
to~(\ref{eq:W_M_Z2}) tests the interpretation that the drop-off of
$W_M(R)$ is due to Z($N_c$) vortices, without requiring gauge fixing
of the SU($N_c$) links~\cite{Stack:2000he}.

The field in the forward light cone immediately after a
collision~\cite{Kovner:1995ja}, at proper time
$\tau\equiv\sqrt{t^2-z^2}\to 0$, is given in light cone gauge by
$A^i=\alpha_1^i + \alpha_2^i$. In turn, before the collision the
individual fields of projectile and target are 2d pure gauges,
\be \label{eq:alphai} 
\alpha^i_m = \frac{i}{g} \, U_m \, \partial^i
U_m^\dagger~~~~,~~~~ \partial^i \alpha^i_m = g \rho_m~,
\ee
where $m=1,\, 2$ labels projectile and target, respectively, and $U_m$
are SU(N) gauge fields.
\begin{figure}[htb]
\begin{center}
\includegraphics[angle=-90,width=0.48\textwidth]{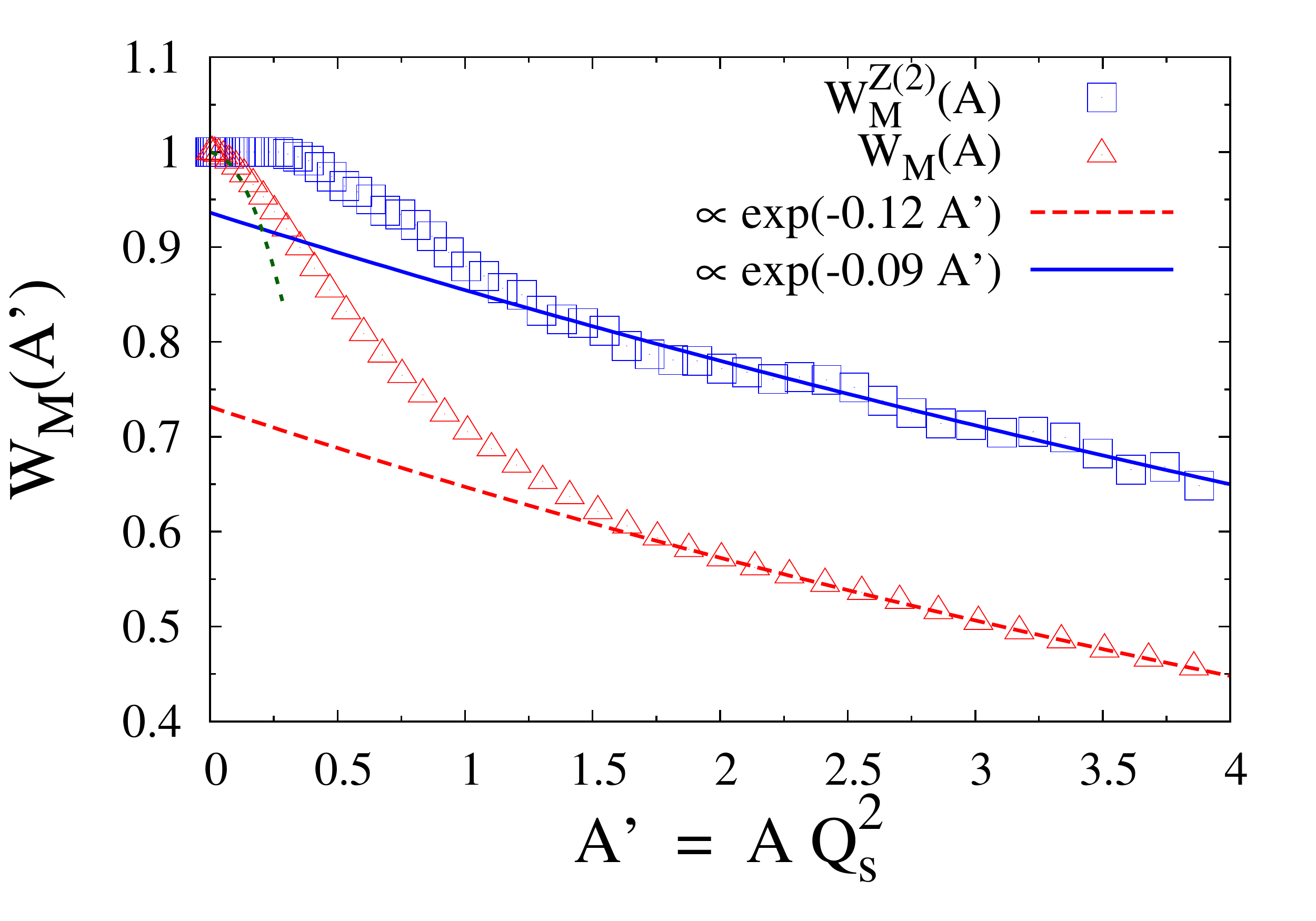}
\end{center}
\vspace*{-0.6cm}
\caption[a]{Expectation value of the magnetic flux loop right after a
  collision of two nuclei (time $\tau=+0$) as a function of its area
  $A'\equiv A\, Q_s^2$. We define $Q_s^2 = (C_F/2\pi)\,g^4
  \mu^2$. Symbols show numerical results for $SU(2)$ Yang-Mills on a $4096^2$
  lattice; the lattice spacing is set by $g^2\mu_L=0.0661$.
  The solid and dashed lines represent fits over the range $4\ge A'\ge
  2$. The short dotted line shows $\cos 2A'$ for $A'<0.3$.}
\label{fig:mfl_tau=0}
\end{figure}
Eqs.~(\ref{eq:alphai}) can be solved either analytically in an
expansion in the field strength~\cite{Kovner:1995ja} or numerically on
a lattice~\cite{CYMlatt}.

The large-$x$ valence charge density $\rho$ is a random
variable\footnote{In the context of large vs.\ small $x$ this variable
denotes the light-cone momentum of a parton relative to the beam
hadron and should not be confused with a transverse coordinate.}. 
For a large nucleus, the effective action describing color charge
fluctuations is quadratic,
\be \label{eq:S2}
S_{\rm eff}[\rho^a] = \frac{\rho^a({\bf x}) \rho^a({\bf
    x})}{2\mu^2}~~~,~~~ \langle \rho^a({\bf x})\, \rho^b({\bf
  y})\rangle =\mu^2 \delta^{ab} \delta({\bf x}-{\bf y})~, 
\ee 
with $\mu^2$ proportional to the thickness of a given
nucleus~\cite{MV}. The variance of color charge fluctuations
determines the saturation scale $Q_s^2 \sim g^4
\mu^2$~\cite{JalilianMarian:1996xn}. The coarse-grained effective
action~(\ref{eq:S2}) applies to (transverse) area elements containing
a large number of large-$x$ ``valence'' charges, $\Delta A_\perp\,
\mu^2 \sim \Delta A_\perp\,Q_s^2/g^4 \gg 1$. The densities
$\rho^a({\bf x})$ at two different points are independent so that their
correlation length within the effective theory is zero. However, this
is not so for the gauge fields $A^i$ which do exhibit a finite
screening length~\cite{m_screen}.

In fig.~\ref{fig:mfl_tau=0} we show numerical results for $W_M$
immediately after a collision. It exhibits area law behavior for loops
larger than $A \gsim 2/Q_s^2$. The corresponding ``magnetic string
tension'' is $\sigma_M/Q_s^2=0.12(1)$. The area law indicates
uncorrelated magnetic flux fluctuations through the Wilson loop and
that the area of magnetic vortices is rather small, their radius being
on the order of $R_{\rm vtx} \sim 0.8/Q_s$. We do not observe a
breakdown of the area law up to $A\sim 4/Q_s^2$, implying that vortex
correlations are small at such distance scales. Also, restricting
to the Z(2) part reduces the magnetic flux through small loops but
$\sigma_M$ is comparable to the full SU(2) result, if somewhat
smaller.

The numerically small vortex size that we find is {\em parametrically}
consistent with the classical Gaussian approximation at weak coupling
which, as already mentioned above, applies for areas $\Delta
A_\perp\gg g^4/Q_s^2$. Corrections to $S_{\rm eff}$ of higher order in
$\rho$~\cite{Srho} as well as due to quantum
fluctuations~\cite{Berges:2012cj} of the fields should be investigated
in the future.

Since the field $\alpha^i_m$ of a single nucleus is a pure gauge it
follows that $W_M^{\rm sngl}(\alpha_m^i)=1$. However, beyond linear
order the field in the forward light cone, $A^i=\alpha^i_1 +
\alpha^i_2$ is not a pure gauge and so $W_M(A^i)\neq
1$.  As discussed in the appendix, for small loops the correction
\be
W_M(\alpha_1^i+\alpha_2^i) - 1 \sim - A^2  \label{eq:W_M_pert}
\ee
is proportional to the square of the area of the loop. There is no
contribution at order $\sim A$, hence no area law behavior and no
vortices, even though there is, of course, a non-zero longitudinal
magnetic field even at the ``naive'' perturbative level:
\be \label{eq:BzCorr}
\frac{g^2}{N_c} \langle \tr B_z({\bf r}) \, B_z({\bf r'})\rangle =
4 \frac{N_c^2}{N_c^2-1} \,Q_{s1}^2 Q_{s2}^2 \log^2 \frac{1}
{|{\bf r} - {\bf r'}|\, \Lambda}~.
\ee

\begin{figure}[htb]
\begin{center}
\includegraphics[angle=-90,width=0.48\textwidth]{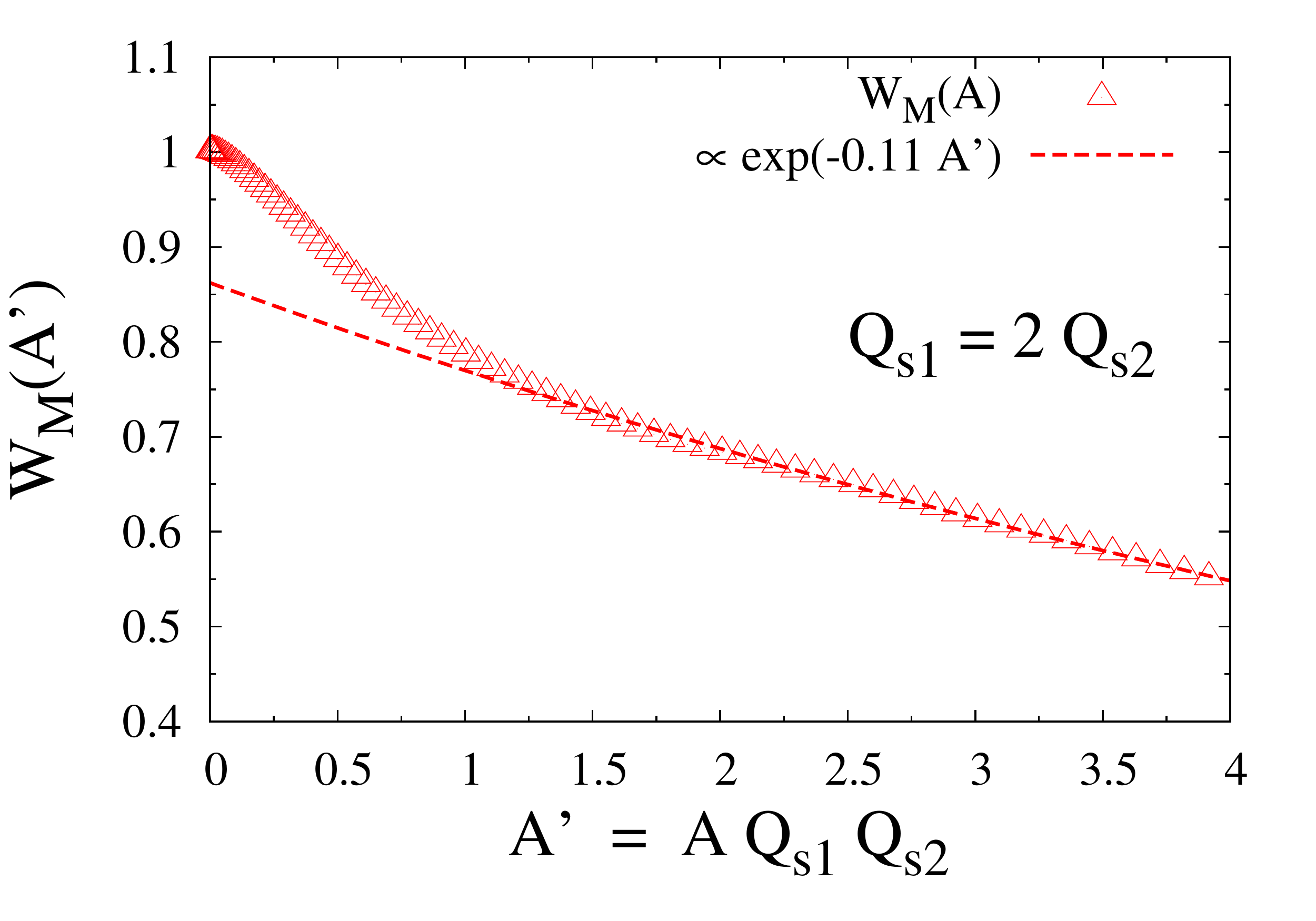}
\end{center}
\vspace*{-0.4cm}
\caption[a]{Same as fig.~\ref{fig:mfl_tau=0} for asymmetric projectile
and target saturation momenta.}
\label{fig:mfl_asym}
\end{figure}
Eq.~(\ref{eq:W_M_pert}) applies for small $AQ_s^2\ll1$ while the
non-perturbative lattice result exhibits area law behavior at
$\tau=+0$ for $AQ_s^2\gsim1$. It indicates the presence of
resummed screening corrections for magnetic fields~\cite{m_screen}. To
see this more explicitly it is useful to notice that $\sigma_M \sim
Q_s^2$ is in fact $\sigma_M \sim Q_{s1} Q_{s2}$, proportional to the
product of single powers of the respective saturation scales of
projectile and target. We have verified this numerically in
fig.~\ref{fig:mfl_asym}. Naive perturbation theory can only produce
even powers of the two-point function $\sim Q_s^2$.

To estimate the density of vortices one can consider a simple
combinatorial model whereby the area $A$ of the loop is covered by
patches of size $1/Q_s^2$ containing a Z(2) vortex with probability $p$.
Averaging over random, uncorrelated vortex fluctuations leads
to~\cite{preskill}
\be
W_M(A) \sim \exp\left( - \frac{\pi^2}{4} \; p(1-p) A Q_s^2\right)~,
\ee
or $\sigma_M = (\pi^2/4) \; p (1-p) Q_s^2$. From this relation we
estimate that the probability of finding a vortex within an area
$1/Q_s^2$ is $p\simeq 1/20$.

\begin{figure}[htb]
\begin{center}
\includegraphics[angle=-90,width=0.48\textwidth]{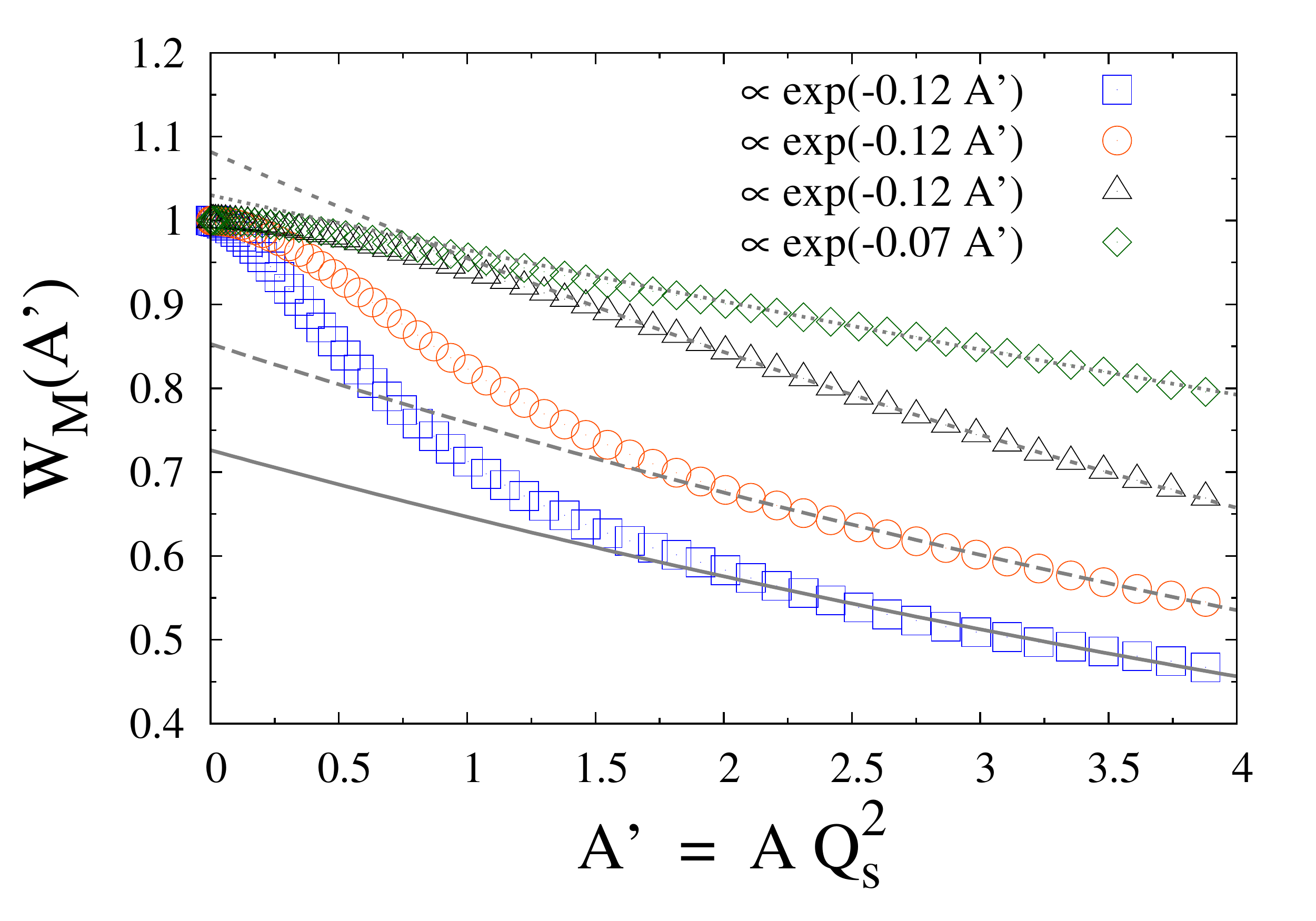}
\end{center}
\vspace*{-0.4cm}
\caption[a]{Time evolution of the magnetic flux loop after a collision
  of two nuclei ($4096^2$ lattice, $g^2\mu_L=0.05$). From bottom to
  top, the curves correspond to time $\tau\times g^2\mu=0$, 1, 2, 3,
  where $g^2\mu \simeq 3Q_s$ so that $\tau=3/(g^2\mu)$ corresponds to
  about $\tau\simeq 1/Q_s$ in physical units.}
\label{fig:mfl_tau>0}
\end{figure}
In fig.~\ref{fig:mfl_tau>0} we show the time evolution of the magnetic
flux loop after a collision. The magnetic field strength decreases due to
longitudinal expansion and so $W_M$ approaches unity. On the other
hand, the onset of area law behavior is pushed to smaller loops,
implying that the size of elementary flux excitations or ``vortices''
decreases; by the time $\tau\sim1/Q_s$ area law behavior is satisfied
even for rather small loops. Since long wavelength magnetic fields
remain even at times $\sim 1/Q_s$, it will be important in the future
to understand the transition of $W_M$ to behavior expected in {\em
  thermal} QCD where $\sigma_M \sim (g^2T)^2$~\cite{kajantie}. In the
context of late-time behavior much beyond $t\sim 1/Q_s$ we refer to
ref.~\cite{Berges} where area law scaling of spatial loops has been
observed for classical field configurations emerging from unstable plasma
evolution. 

We have also investigated the dependence of the magnetic flux loop in
the adjoint representation on its area,
\be \label{eq:adjLoop}
W_M^{\rm adj} = \frac{1}{N_c^2-1} \left< |\tr M|^2 -1\right>~,
\ee
and found behavior similar to fig.~\ref{fig:mfl_tau>0}. The adjoint
magnetic string tension is about two times larger, as expected
from~(\ref{eq:adjLoop}).

\begin{figure}[htb]
\begin{center}
\hspace*{-2.2cm}\includegraphics[angle=-90,width=0.7\textwidth]{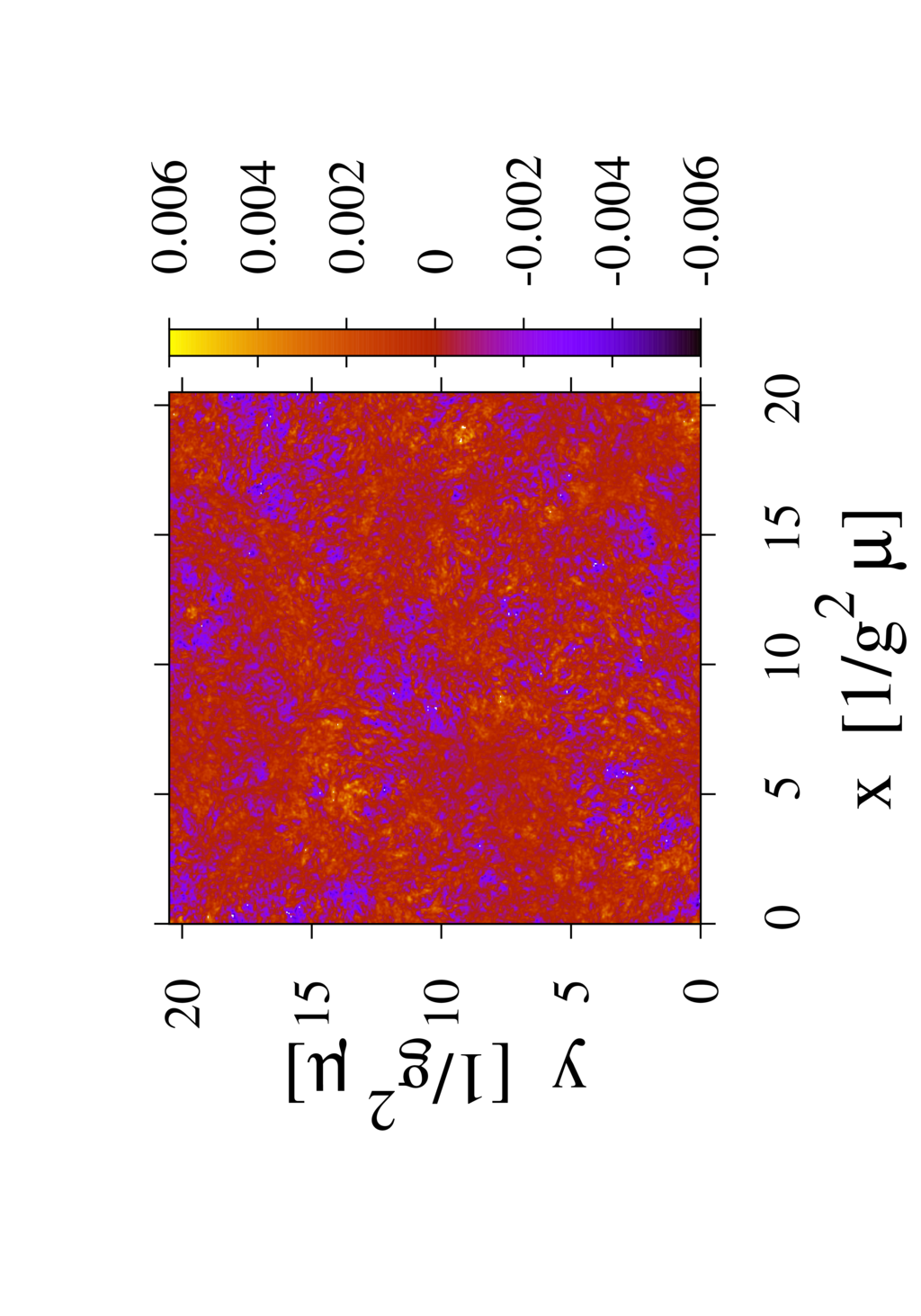}
\hspace*{-2.2cm}\includegraphics[angle=-90,width=0.7\textwidth]{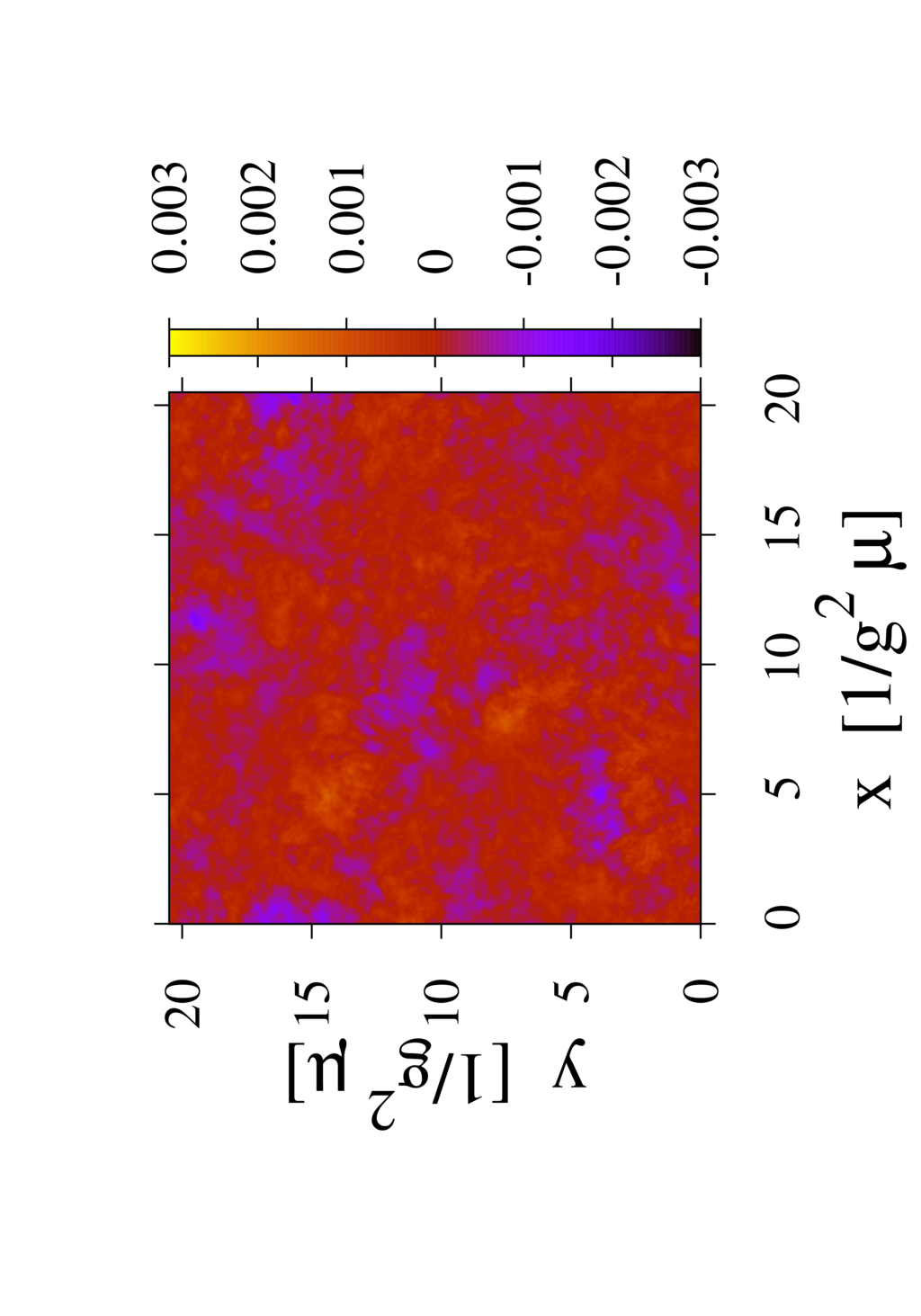}
\end{center}
\vspace*{-0.4cm}
\caption[a]{Color-3 component of the magnetic field $F_{xy}^3(x,y)$ in
  the transverse plane at time $\tau=+0$ (top) and $1/g^2\mu\sim
  1/3Q_s$ (bottom) for a single configuration of color charge sources
  $\rho$.}
\label{fig:Bz}
\end{figure}
The third color component of the longitudinal magnetic field is shown
in fig.~\ref{fig:Bz}, using a random residual gauge for $A^i$. Domain-like
structures where the magnetic field is either positive or negative are
clearly visible; they lead to the above-mentioned area law of the
Wilson loop. Also, one can see that in time the magnetic fields become
weaker and smoother.

\begin{figure}[htb]
\begin{center}
\includegraphics[width=0.48\textwidth]{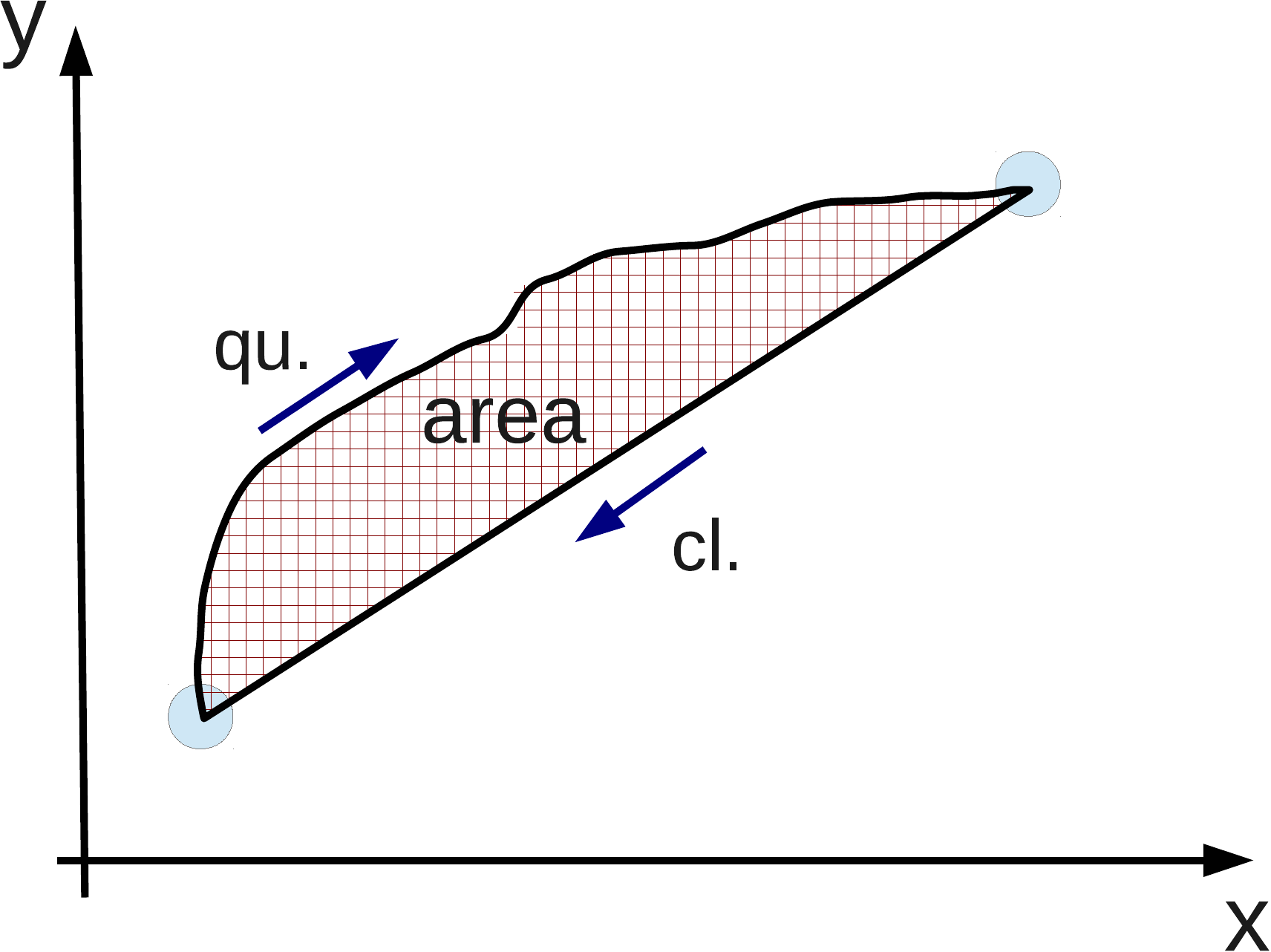}
\end{center}
\vspace*{-0.7cm}
\caption[a]{Area enclosed by a quantum mechanical path shifted by
  about one de~Broglie wavelength from the extremal classical path.}
\label{fig:paths}
\end{figure}
Thus far we have not addressed the {\em longitudinal} structure of the
initial fields. Our solution of eqs.~(\ref{eq:alphai}) is boost
invariant and so, naively, the two-dimensional vortex structures
mentioned above would form boost invariant strings. However, this
simple picture could be modified by longitudinal smearing of the
valence charge distributions~\cite{Fukushima:2007ki} and therefore
requires more detailed consideration.

The magnetic fields modify the propagation of semi-hard modes with
$p_T$ not too far above $Q_s$. Quantum mechanically, the
transition amplitude from a state $\ket{x_i,t_i}$ to $\ket{x_f,t_f}$
is given by a Feynman sum over paths,
\bea
& & \int_0^\infty ds \int {\cal D}x^\mu
\left< \exp\,
i \int\limits_0^s  d\tau\left(m\dot{x}^2 + g A_\mu \dot{x}^\mu\right)\right>
\sim \nonumber\\
& & \int_0^\infty ds \int {\cal D}x^\mu \exp\,
\left( i \int\limits_0^s d\tau\, m\dot{x}^2\right)\;
\exp(-\sigma_MA)~,   \label{eq:PathIntArea}
\eea
where $x^\mu(\tau)$ is a parametrization of the path with the given
boundary conditions and length $s$; and $\dot{x}^\mu = dx^\mu/d\tau$.
Here, the area $A$ is that enclosed by a quantum mechanical path from
the initial to the final point returning to ${\bf x}_i$ via the
classical path; see fig.~\ref{fig:paths}. 
The classical path is obtained by extremizing the action but a single
path is a set of measure zero. Semi-classical paths can dominate the
integral only if there is constructive interference among neighboring
paths from within a de~Broglie distance. On the other hand,
destructive interference of such paths leads to Anderson localization of
the wave function.

Hence, up to a numerical factor, the area in
eq.~(\ref{eq:PathIntArea}) should be given by $A\sim
s/p_T$. Integrating over the Schwinger parameter then leads to the
propagator
\be
\frac{i}{p^2 +i \sigma_M \frac{m}{p_T}}~,
\ee
where $\sigma_M =0.12\, Q_s^2$ from above and $m$ is the mass
(time-like virtuality) of the particle. This expression accounts
for corrections to free propagation and could be useful for studies of
the dynamics of the very early stage of a heavy-ion collision.

We obtain a rather interesting picture of the very early stage of
ultrarelativistic heavy-ion collisions. Magnetic Wilson loops of area
$\gsim 2/Q_s^2$ effectively exhibit area law behavior which implies
uncorrelated magnetic Z(N) vortex-like flux beyond the scale $R_{\rm
  vtx}\sim 0.8/Q_s$. We do expect that corrections
to this picture appear at much larger distance scales and we intend to
study these in detail in the future. The vortex structure of the
longitudinal magnetic field modifies propagation of particle-like
modes with de~Broglie wavelength somewhat larger than $Q_s$.

\vspace*{1cm}
\begin{acknowledgments}
We thank A.~Kovner, L.~McLerran, P.~Orland and R.~Pisarski for helpful
comments.  A.D.\ and E.P.\ gratefully acknowledge support by the DOE
Office of Nuclear Physics through Grant No.\ DE-FG02-09ER41620; and
from The City University of New York through the PSC-CUNY Research
grant 66514-00~44.

\end{acknowledgments}

\appendix

\section{Appendix: Perturbative limit of the magnetic Wilson loop
  at $\tau=0$}

In this appendix we outline the ``naive'' perturbative expansion of
the loop with the Gaussian contractions. We stress that since magnetic
fields at $\tau=0$ are screened over distances $\sim 1/(5
Q_s)$~\cite{m_screen}, that this naive expansion can not be applied in
the regime of interest in the present paper.

To determine $W_M(A^i)$ we need to determine the ``correction'' in
$A^i=\alpha_1^i + \alpha_2^i$ from a pure gauge. From the
Baker-Campbell-Hausdorff relation,
\be \label{eq:exp_com}
W_M(A^i) \simeq \frac{1}{N_c} \tr \exp\left( - \frac{1}{2} [X_1,X_2]\right)~,
\ee
where terms of third order in the fields have been dropped from the
exponent; and $X_m$ is $\alpha_m^i$ integrated along the loop.

In the weak field limit~\cite{Kovner:1995ja}
\be \label{eq:alpha_Phi}
\alpha_m^i = -\partial^i\Phi_m + \frac{ig}{2}\left(
\delta^{ij} - \partial^i \frac{1}{\nabla_\perp^2}\partial^j\right)
\left[\Phi_m,\partial^j\Phi_m\right] + \cdots
\ee
The first term on the rhs of eq.~(\ref{eq:alpha_Phi}) does not
contribute to the integral of $\alpha_m^i$ over a closed loop.

We can express the square of the exponent on the
r.h.s.\ of~(\ref{eq:exp_com}) as
\be \label{eq:h2}
h^2 \equiv \frac{1}{16} f^{abc} f^{dec} X_1^a X_1^d X_2^b X_2^e~,
\ee
so that for two colors
\be
W_M(A^i) \simeq \left< \cos h\right>_{\rho_1,\rho_2}
\simeq 1 - \frac{1}{2} \langle h^2 \rangle~.
\ee
Eq.~(\ref{eq:h2}) contains four integrations over the periphery of the
loop, two of which will be removed by the
$\langle\cdot\rangle_{\rho_1}$ and $\langle\cdot\rangle_{\rho_2}$
contractions; c.f.\ eq.~(\ref{eq:S2}). Hence $\langle h^2\rangle \sim
A^2 \mu^4$ is proportional to the square of the area of the
loop. Fig.~\ref{fig:mfl_tau=0} shows that by matching to the lattice
data at $A'\ll1$ we estimate $h\approx 2 A'$.


\end{document}